\documentclass[letterpaper, 10 pt, conference]{ieeeconf}  

\IEEEoverridecommandlockouts                              
\usepackage[utf8]{inputenc}
\usepackage[T1]{fontenc}
\usepackage{graphicx}
\usepackage{amsmath}

\usepackage{amssymb}  

\title{\LARGE \bf
Hydrodynamic Whispering: Enabling Near-Field Silent Communication via Artificial Lateral Line Arrays
}

\author{Yuan-Jie Chen$^{1}$
\thanks{$^{1}$State Key Laboratory of Ocean Sensing \& Ocean College, Zhejiang University, Zhoushan, 316021, China (Email: chenyuanjie201@gmail.com)}%
}

\begin{document}

\maketitle
\thispagestyle{empty}
\pagestyle{empty}

\begin{abstract}

To address the imperative for covert underwater swarm coordination, this paper introduces "Hydrodynamic Whispering," a near-field silent communication paradigm utilizing Artificial Lateral Line (ALL) arrays. Grounded in potential flow theory, we model the transmitter as an oscillating dipole source. The resulting pressure field exhibits steep near-field attenuation (scaling with $1/r^2$), naturally delimiting a secure "communication bubble" with intrinsic Low Probability of Interception (LPI) properties. We propose a transceiver architecture featuring a Binary Phase Shift Keying (BPSK) modulation scheme adapted for mechanical actuator inertia, coupled with a bio-inspired 24-sensor conformal array. To mitigate low Signal-to-Noise Ratio (SNR) in turbulent environments, a Spatio-Temporal Joint Processing framework incorporating Spatial Matched-Field Beamforming is developed. Simulation results demonstrate that the system achieves an array gain of approximately 13.8 dB and maintains a near-zero Bit Error Rate (BER) within the effective range. This study validates the feasibility of utilizing localized hydrodynamic pressure fluctuations for reliable and secure short-range underwater networking.

\end{abstract}

\section{INTRODUCTION}

With increasing demands for marine exploitation and underwater tactical operations, Autonomous Underwater Vehicle (AUV) swarms have emerged as a key force for collaborative sensing, mine countermeasures, and distributed attacks \cite{c1}. In such cluster operations, high-frequency, short-range information exchange between individuals is critical. Currently, the main underwater communication is based on hydroacoustic and underwater optical wireless communication (UOWC) \cite{c2}. However, these modalities face significant limitations in tactical scenarios: acoustic waves, while capable of long-range transmission, suffer from severe multipath latency \cite{c3}, and their omnidirectional radiation compromises stealth by exposing the source's location. In contrast, optical communication, despite its high bandwidth, is hypersensitive to water turbidity and requires strict Line-of-Sight (LOS) alignment \cite{c4}. Consequently, developing a novel communication modality that is short-range, highly reliable, and characterized by electromagnetic and acoustic silence has become an urgent scientific imperative in the field of underwater robotics.

Nature offers an elegant solution to this engineering challenge. The lateral line system evolved by fish, often referred to as a "tactile eye," enables the perception of minute pressure variations and velocity gradients in the surrounding flow \cite{c5}. Biological studies reveal that in high-density schooling, individuals do not rely solely on vision, but maintain position and synchronization by sensing the wake fields generated by neighbors' tail beats via their lateral lines \cite{c6}, \cite{c7}. This interaction, mediated by the fluid medium, represents a quintessential mechanism of short-range, silent hydrodynamic communication. Despite this biological precedent, current research on ALL remains predominantly confined to "passive perception" domains \cite{c8}, such as dipole source localization \cite{c9}, obstacle identification \cite{c10}, and flow velocity measurement, with scant literature exploring their potential for "active communication."

Bridging this research gap, this paper proposes a novel hydrodynamic communication paradigm that uses ALL for active information transmission. We first establish a rigorous physical channel model based on potential flow theory \cite{c11}, deriving the near-field pressure distribution of an oscillating sphere in viscous fluid to quantify the nonlinear attenuation relationship between communication range and signal intensity. Then a complete transmitter-receiver architecture is presented, incorporating a BPSK modulation strategy adapted for mechanical inertia and a streamlined array layout. To overcome the low SNR inherent in single-sensor reception, we innovatively introduce array signal processing techniques from the radar and sonar domains, designing a spatially matched filter tailored to the characteristics of the dipole field \cite{c12}. Comprehensive simulations of the system verify the feasibility of the proposed scheme through the analysis of eye diagrams, beamforming gains, and BER performance.

The remainder of this paper is organized as follows: Section 2 details the system modeling and physical channel properties; Section 3 describes the proposed modulation and spatio-temporal signal processing algorithms; Section 4 presents the numerical simulation results and performance analysis; and Section 5 concludes the paper with a discussion on future work.

\begin{figure}[thpb]
   \centering
   \includegraphics[width=0.9\columnwidth]{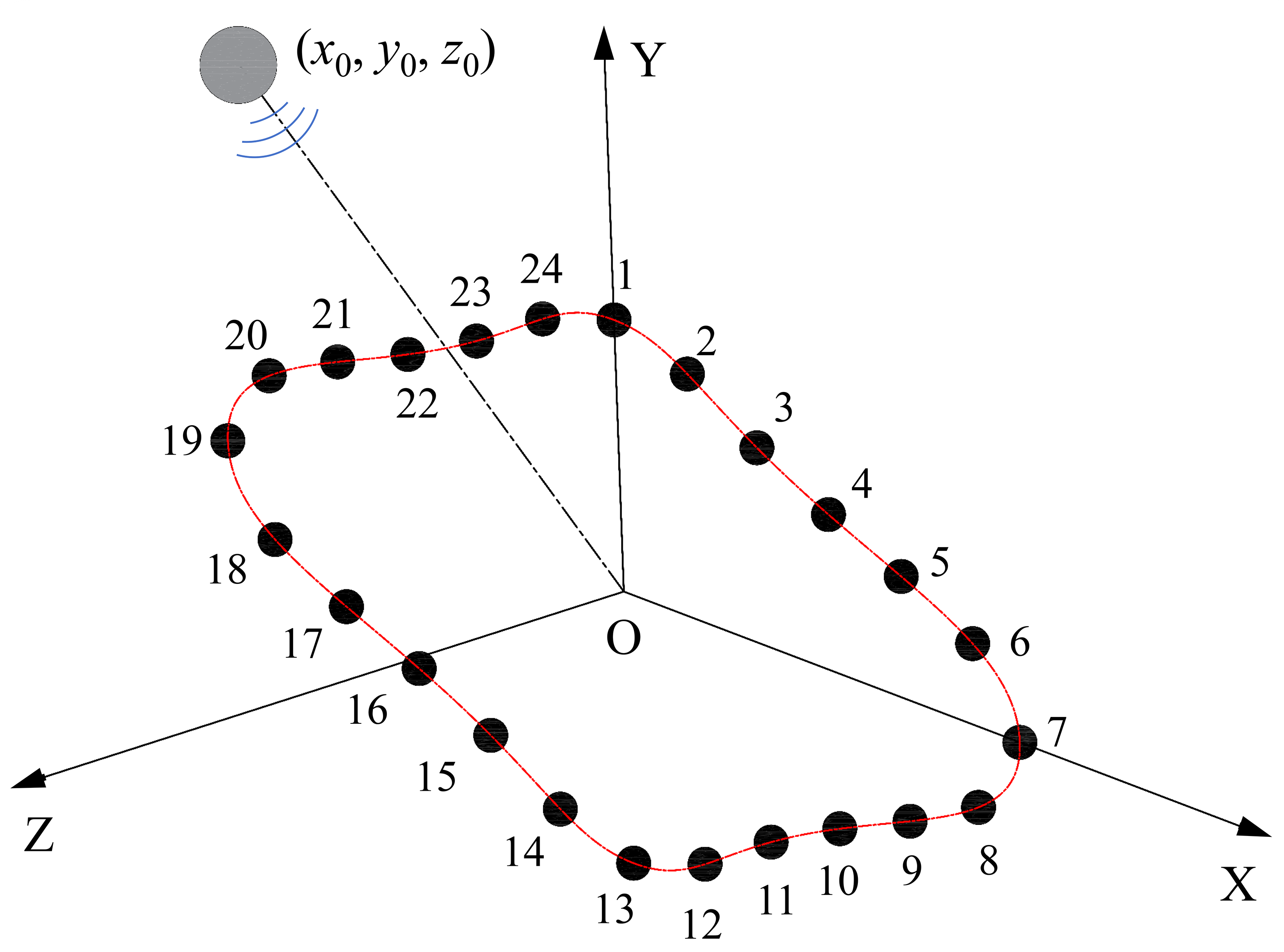}
   \caption{Schematic of the submerged hydrodynamic communication scenario and coordinate system.}
   \label{fig1_lb}
\end{figure}

\section{System Modeling and Physical Principles}

\subsection{Problem Setup and Coordinate System}

As illustrated in Fig. \ref{fig1_lb}, we consider a submerged communication scenario in an unbounded, incompressible, and inviscid fluid domain with density $\rho$. The system consists of a transmitter (a vibrating spherical source) and a receiver (an artificial lateral line array).
A Cartesian coordinate system is established with the origin $\mathbf{O}$ located at the geometric center of the receiver array. The transmitter is positioned at $\mathbf{r}_0 = [x_0, y_0, z_0]^T$. The position vector of an arbitrary point in the flow field is indicated as $\mathbf{r} = [x, y, z]^T$, and the Euclidean distance from the source to this arbitrary point is defined as $d = \|\mathbf{r} - \mathbf{r}_0\|$.

\subsection{Transmitter Modeling: The Dipole Source}

The transmitter is modeled as a rigid sphere of radius $a$, oscillating along the $x$-axis. This mechanical oscillation perturbs the surrounding fluid, generating a non-steady flow field. Given the small scale of the sphere and the frequency range of interest (low Reynolds number), the generated flow field can be accurately approximated using Potential Flow Theory.Let the displacement of the sphere center be 
\begin{equation}
    x(t) = A \sin(\omega t), 
\end{equation}
where $A$ is the vibration amplitude and $\omega = 2\pi f_c$ is the angular carrier frequency. The instantaneous velocity of the sphere is 
\begin{equation}
    U(t) = \dot{d}(t) = A\omega \cos(\omega t).
\end{equation}

According to potential flow theory, a moving sphere in an infinite fluid is equivalent to a Dipole Source. The velocity potential $\phi(\mathbf{r}, t)$ at position $\mathbf{r}$ is given by:
\begin{equation}
    \phi(\mathbf{r}, t) = -\frac{a^3}{2r^2} U(t) \cos\theta,
\end{equation}
where $\theta$ is the polar angle with respect to the vibration axis ($x$-axis), and $\cos\theta = x/r$.The dynamic pressure distribution $p(\mathbf{r}, t)$ can be derived from the unsteady Bernoulli equation (linearized for small perturbations):
\begin{equation}
    p(\mathbf{r}, t) = -\rho \frac{\partial \phi(\mathbf{r}, t)}{\partial t}.
\end{equation}

Substituting $\phi(\mathbf{r}, t)$ and the acceleration $\dot{U}(t) = -A\omega^2 \sin(\omega t)$ into the equation, we obtain the governing equation for the pressure field:
\begin{equation}
    p(\mathbf{r}, t) = \frac{\rho a^3}{2r^2} \cos\theta \cdot \dot{U}(t).
\end{equation}

Substituting the explicit time-domain expression for acceleration, the final pressure field model is:
\begin{equation}
    p(x, y, z, t) = -\frac{\rho A \omega^2 a^3}{2} \cdot \underbrace{\frac{x}{(x^2 + y^2 + z^2)^{3/2}}}_{\text{Spatial Term } G(\mathbf{r})} \cdot \sin(\omega t).
\end{equation}

Eq. (6) elucidates the intrinsic physics of the hydrodynamic channel: the pressure amplitude scales with the source acceleration ($A\omega^2$), while the spatial distribution exhibits a steep $1/r^2$ near-field attenuation—distinct from the $1/r$ far-field acoustic decay—thereby creating a secure "silent bubble" with LPI. Additionally, the angular dependence introduces a characteristic "figure-8" dipole directivity.

\subsection{Channel Characteristics and Noise Model}

Unlike ideal free-space propagation, the underwater hydrodynamic channel is subject to background flow noise. The dominant interference in the near-field regime arises from ambient turbulence.The noise pressure $n(t)$ sensed by the receiver is modeled as a stochastic process. The Power Spectral Density (PSD) of turbulent pressure fluctuations typically follows the Kolmogorov's -5/3 power law in the inertial subrange:
\begin{equation}
    S_n(f) \propto f^{-5/3}.
\end{equation}

This implies that background noise is predominantly concentrated in the low-frequency band. Consequently, selecting a carrier frequency $f_c$ (e.g., 40 Hz) sufficiently higher than the dominant turbulence frequencies allows for effective spectral separation, improving the SNR.

\subsection{Receiver Modeling: The Artificial Lateral Line Array}
The receiver is an ALL system consisting of $N$ discrete pressure sensors distributed on the AUV hull. Let the position of the $i$-th sensor be $\mathbf{m}_i = [x_i, y_i, z_i]^T$.The measured signal $y_i(t)$ at the $i$-th sensor is the superposition of the transmitted hydrodynamic signal and the local environmental noise:
\begin{equation}
    y_i(t) = p(\mathbf{m}_i, t) + n_i(t).
\end{equation}

Substituting the dipole model from Eq. (6), we can express the array output in vector form:
\begin{equation}
    \mathbf{y}(t) = \mathbf{h} \cdot s(t) + \mathbf{n}(t).
\end{equation}
where $\mathbf{y}(t) = [y_1(t), \dots, y_N(t)]^T$ is the $N \times 1$ received signal vector.$s(t) = -\frac{\rho A \omega^2 a^3}{2} \sin(\omega t)$ encapsulates the source strength and time-varying modulation.$\mathbf{h} = [G(\mathbf{m}_1), \dots, G(\mathbf{m}_N)]^T$ is the spatial fingerprint vector, representing the theoretical geometric response of the array to the dipole source.$\mathbf{n}(t)$ is the additive Gaussian noise vector (approximated) with covariance matrix $\sigma_n^2 \mathbf{I}$.This vector formulation (Eq. 9) forms the mathematical basis for the spatial beamforming algorithms discussed in Section 3.

\section{Proposed Communication Scheme}

Based on the physical channel model established in Section 2, we propose a comprehensive hydrodynamic communication framework. This section details the transmitter modulation strategy adapted for mechanical actuators and the two-stage receiver architecture: Spatial Matched-Field Beamforming followed by Temporal Coherent Demodulation.

\subsection{Transmitter: Mechanical BPSK Modulation} 
Unlike electromagnetic antennas that can switch frequencies instantaneously, hydrodynamic sources (e.g., oscillating spheres or vibrating fins) are constrained by mechanical inertia. To maximize bandwidth efficiency while respecting these physical constraints, we adopt Binary Phase Shift Keying (BPSK).Let the binary data stream be denoted as $\{b_k\}$, where $b_k \in \{0, 1\}$. The $k$-th symbol is mapped to a polarity coefficient 
$d_k$:
\begin{equation}
    d_k = \begin{cases} 
+1, & \text{if } b_k = 1 \text{ (Phase } 0) \\
-1, & \text{if } b_k = 0 \text{ (Phase } \pi)
\end{cases}.
\end{equation}

The modulated source vibration signal $s_{mod}(t)$ for the $k$-th symbol period $t \in [(k-1)T_s, kT_s]$ is defined as:
\begin{equation}
    s_{mod}(t) = A \cdot d_k \cdot \sin(\omega_c t),
\end{equation}
where $T_s = 1/R_b$ is the symbol duration.Remark on Mechanical Feasibility: The phase shift of $\pi$ physically corresponds to inverting the vibration direction. For a reciprocating motor or voice-coil actuator, this transition can be smoothed by a shaping filter to prevent mechanical shock, though for theoretical derivation, we assume ideal switching.

\subsection{Stage 1: Spatial Matched-Field Beamforming}
The primary challenge in hydrodynamic communication is the low SNR at individual sensors and the spatial non-uniformity of the dipole field.As derived in Eq. (9), the received signal vector is $\mathbf{y}(t) = \mathbf{h} s_{mod}(t) + \mathbf{n}(t)$.A simple summation of sensor outputs $\sum y_i(t)$ would be detrimental because the dipole field contains both positive and negative pressure zones (the "figure-8" pattern). Sensors on opposite sides of the source would cancel each other out.To address this, we employ a Spatial Matched Filter (also known as a Beamformer). The objective is to design a weight vector $\mathbf{w} \in \mathbb{R}^{N \times 1}$ to maximize the SNR of the combined output $y_{\Sigma}(t)$.

\subsubsection{Optimal Weight Derivation}
According to array signal processing theory, the optimal weight vector that maximizes SNR in spatially white noise is proportional to the steering vector (the spatial fingerprint). Thus, we set:
\begin{equation}
    \mathbf{w} = \mathbf{h} = [G(\mathbf{m}_1), G(\mathbf{m}_2), \dots, G(\mathbf{m}_N)]^T.
\end{equation}

Here, $\mathbf{h}$ is calculated a priori based on the known geometry of the array and the estimated relative position of the source.

\subsubsection{Beamforming Output}
The spatially fused signal $y_{\Sigma}(t)$ is obtained by the inner product:
\begin{equation}
    y_{\Sigma}(t) = \frac{1}{\|\mathbf{h}\|^2} \mathbf{w}^T \mathbf{y}(t) = \frac{1}{\|\mathbf{h}\|^2} \sum_{i=1}^{N} G(\mathbf{m}_i) y_i(t).
\end{equation}

Normalization by $\|\mathbf{h}\|^2$ ensures the signal amplitude remains consistent.Substituting $\mathbf{y}(t)$ into the equation:
\begin{equation}
    y_{\Sigma}(t) = \frac{1}{\|\mathbf{h}\|^2} \left( (\mathbf{h}^T \mathbf{h}) s_{mod}(t) + \mathbf{h}^T \mathbf{n}(t) \right) = s_{mod}(t) + \tilde{n}(t),
\end{equation}
where the term $\mathbf{h}^T \mathbf{h} = \sum G(\mathbf{m}_i)^2$ is always positive. This operation effectively "rectifies" the negative pressure zones, forcing all signal components to add up constructively (Coherent Addition). Since the noise $n_i(t)$ is uncorrelated across sensors, the post-processing noise variance is reduced by a factor of $N$ (Array Gain $\approx 10\log_{10}N$).

\subsection{Stage 2: Temporal Coherent Demodulation}
After spatial processing, the signal $y_{\Sigma}(t)$ is a cleaner, scalar BPSK waveform. The second stage extracts the digital bits using Coherent Demodulation in the time domain.The receiver generates a local reference carrier $c(t) = \sin(\omega_c t)$ synchronized with the transmitter. The decision variable $D_k$ for the $k$-th bit is computed using an Integrate-and-Dump filter:
\begin{equation}
    D_k = \int_{(k-1)T_s}^{kT_s} y_{\Sigma}(t) \cdot c(t) \, dt.
\end{equation}

Substituting $y_{\Sigma}(t) \approx A d_k \sin(\omega_c t)$ (ignoring noise for brevity):
\begin{equation}
    D_k \approx \int_{0}^{T_s} A d_k \sin^2(\omega_c t) \, dt = A d_k \cdot \frac{T_s}{2}
\end{equation}
(Assuming $T_s$ is an integer multiple of the carrier period).

\subsubsection{Decision Rule}
The bit is recovered based on the polarity of $D_k$:
\begin{equation}
    \hat{b}_k = \begin{cases} 
1, & \text{if } D_k > 0 \\
0, & \text{if } D_k \le 0 
\end{cases}
\end{equation}

\section{Numerical Simulation and Performance Analysis}

To evaluate the feasibility and robustness of the proposed hydrodynamic communication system, a high-fidelity numerical simulation platform was constructed using MATLAB. This section presents the simulation setup, visualizes the physical field distribution, analyzes the signal enhancement achieved by the array, and quantifies the BER performance under varying noise conditions.

\subsection{Simulation Setup and Parameters}
The simulation environment models a typical underwater operation scenario. The fluid medium is water with density $\rho = 1000 \, \text{kg/m}^3$. The transmitter is a spherical dipole source oscillating along the $x$-axis. The receiver is a conformal array consisting of $N=24$ pressure sensors arranged in a streamlined dual-line configuration, mimicking the lateral line of a fish.The key system parameters are summarized in Table 1.
\begin{table}[htbp]
\centering
\caption{Key Simulation Parameters}
\label{tab:simulation_parameters}
\begin{tabular}{lll}
\hline
\textbf{Parameter} & \textbf{Symbol} & \textbf{Value} \\
\hline
\multicolumn{3}{l}{\textit{Physical Channel}} \\
Source radius & $a$ & $0.125\,\mathrm{m}$ \\
Source amplitude & $A$ & $15\,\mathrm{mm}$ \\
Carrier frequency & $f_c$ & $40\,\mathrm{Hz}$ \\
\hline
\multicolumn{3}{l}{\textit{Communication}} \\
Bit rate & $R_b$ & $20\,\mathrm{bps}$ \\
Sampling rate & $f_s$ & $2000\,\mathrm{Hz}$ \\
\hline
\multicolumn{3}{l}{\textit{Receiver Array}} \\
Number of sensors & $N$ & $24$ \\
Array geometry & -- & Dual-line \\
Communication distance & $d$ & $70 \mathrm{mm}$ \\
\hline
\end{tabular}
\end{table}

\begin{figure}[thpb]
   \centering
   \includegraphics[width=0.9\columnwidth]{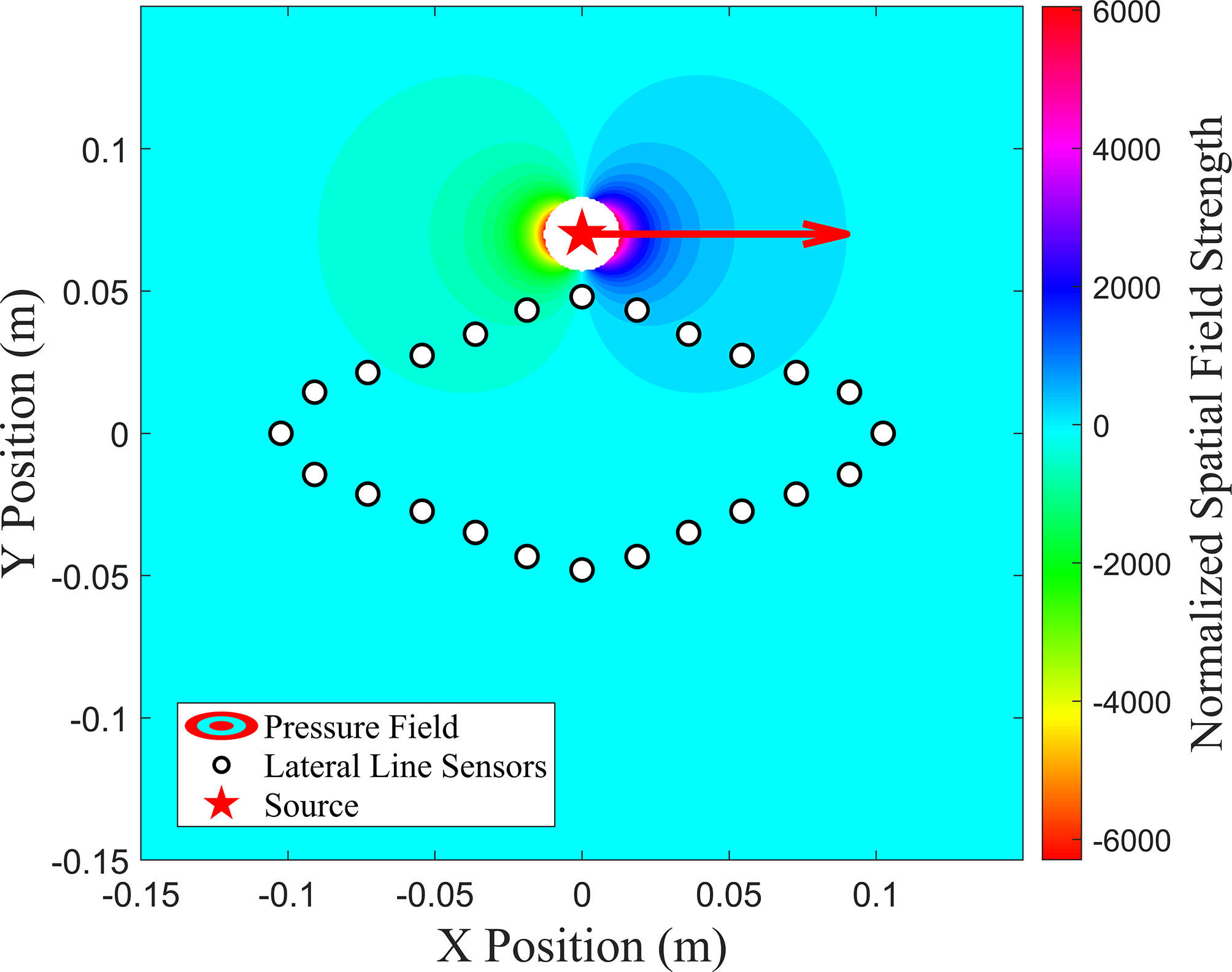}
   \caption{The "Spatial Fingerprint": Dipole Field}
   \label{fig2_lb}
\end{figure}

\subsection{Spatial Field Characterization}
Fig. \ref{fig2_lb} visualizes the normalized pressure field distribution generated by the dipole source.Consistent with the theoretical derivation in Eq. (6), the pressure field exhibits a distinct "figure-8" directional pattern.High-Intensity Zone: The pressure gradient is strongest along the vibration axis ($x$-axis), which is the optimal region for sensor placement.Blind Zone: A zero-pressure nodal plane exists perpendicular to the vibration axis ($y$-axis), where communication is theoretically impossible.The proposed array layout spans the high-intensity region ($x \in [-0.1, 0.1]$), ensuring that the majority of sensors capture effective signal energy. This geometric matching between the array layout and the physical field is crucial for maximizing the input SNR.

\begin{figure*}[thpb]
   \centering
   \includegraphics[width=1.2\columnwidth]{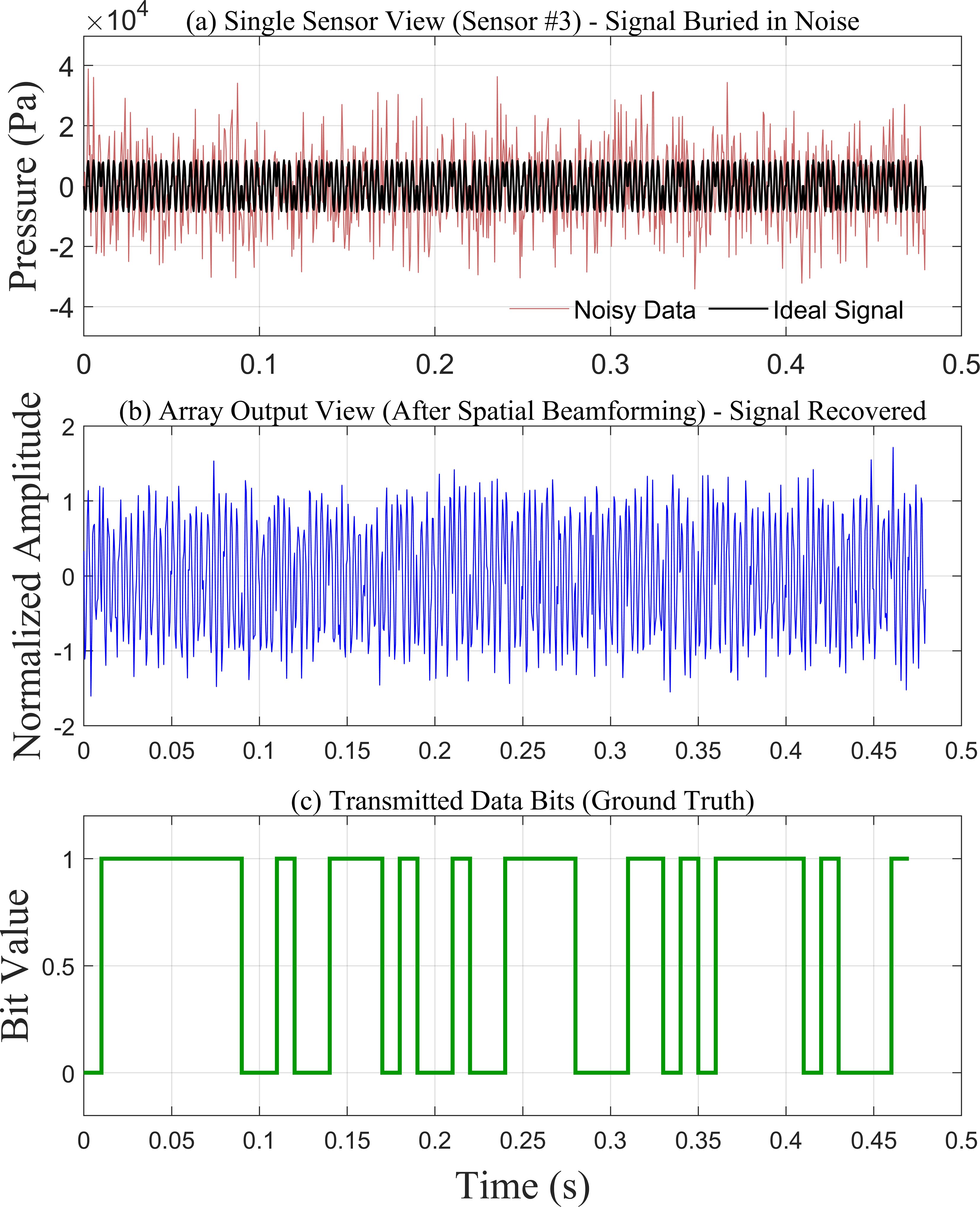}
   \caption{Beamforming Effect}
   \label{fig3_lb}
\end{figure*}

\subsection{Signal Enhancement via Spatial Beamformings}
To demonstrate the efficacy of the proposed spatial matched filter, we compare the time-domain waveforms before and after array processing. We introduce Gaussian white noise to simulate a challenging environment with an input SNR of approximately $-5 \, \text{dB}$ at the single-sensor level.Fig. 2 presents the comparison results:(a) Single Sensor Response: The raw signal from a single sensor (e.g., Sensor \#3) is severely corrupted by noise. The BPSK phase transitions are completely submerged, making direct demodulation impossible.(b) Array Beamforming Output: After applying the spatial matched filter ($\mathbf{w}^T \mathbf{y}$), the signal quality is dramatically improved. The incoherent noise components from 24 sensors average out, while the coherent signal components add constructively. The output waveform clearly reconstructs the BPSK phase shifts (polarity inversions), validating the theoretical Array Gain of $10 \log_{10}(24) \approx 13.8 \, \text{dB}$.

\begin{figure}[thpb]
   \centering
   \includegraphics[width=0.9\columnwidth]{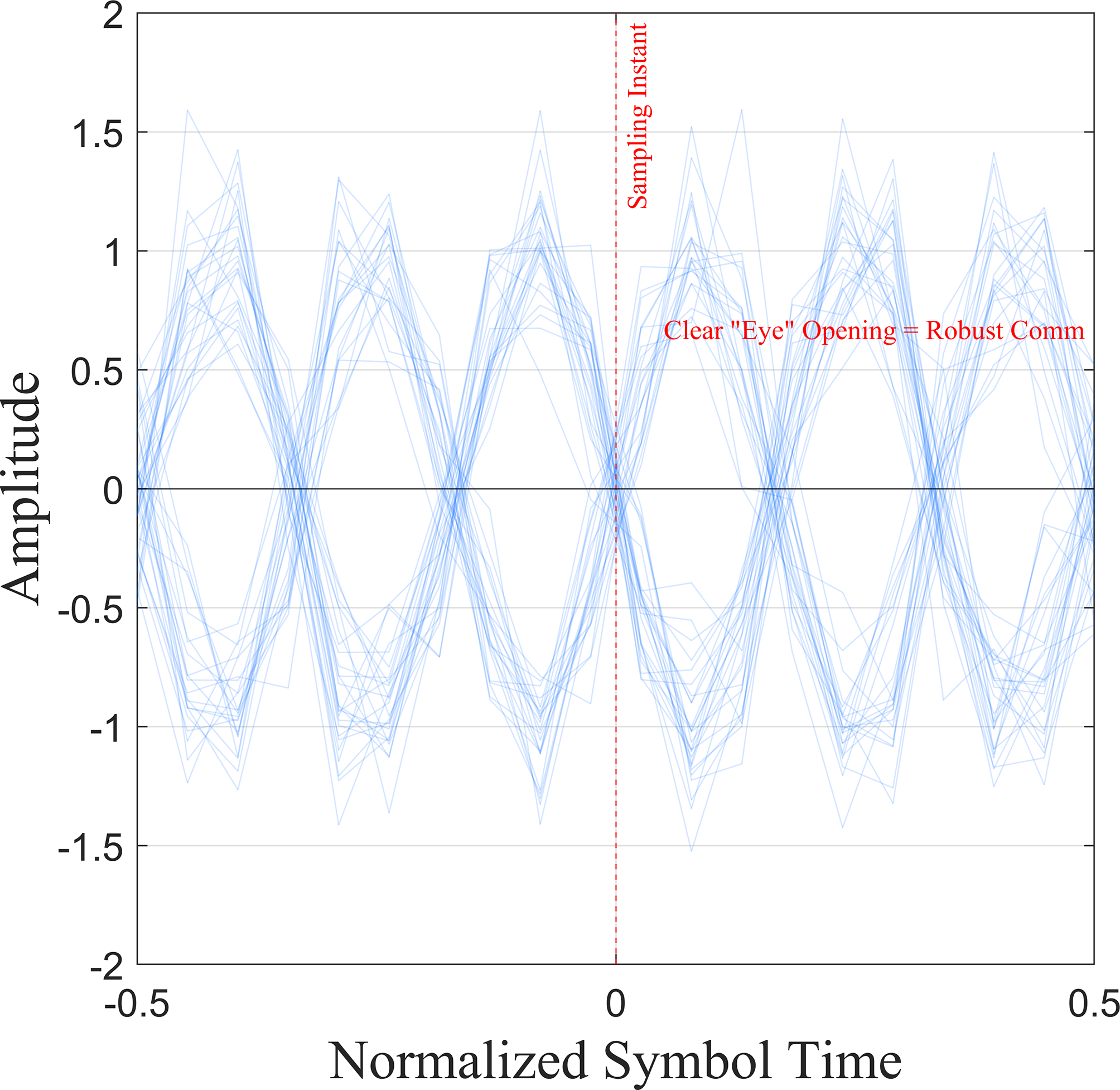}
   \caption{BPSK Eye Diagram}
   \label{fig4_lb}
\end{figure}

\subsection{Demodulation Performance and Eye Diagram}
The quality of the digital communication link is rigorously evaluated using the Eye Diagram, a comprehensive metric for assessing the system's resilience against Inter-Symbol Interference (ISI) and additive noise. As illustrated in Fig. \ref{fig4_lb} for a transmission distance of $d=0.07 \, \text{m}$, the eye pattern manifests critical characteristics that corroborate the effectiveness of the proposed beamforming algorithm. Specifically, the diagram exhibits a wide vertical opening with a distinct separation between the $+1$ and $-1$ decision regions; this substantial noise margin indicates that the spatial matched filter has successfully suppressed incoherent turbulence noise, thereby maximizing the SNR at the decision instant. Furthermore, the zero-crossing points are tightly clustered with negligible temporal jitter, implying that the received signal preserves excellent phase synchronization with the local carrier and minimizes the risk of timing errors during the sampling process.

\begin{figure}[thpb]
   \centering
   \includegraphics[width=0.9\columnwidth]{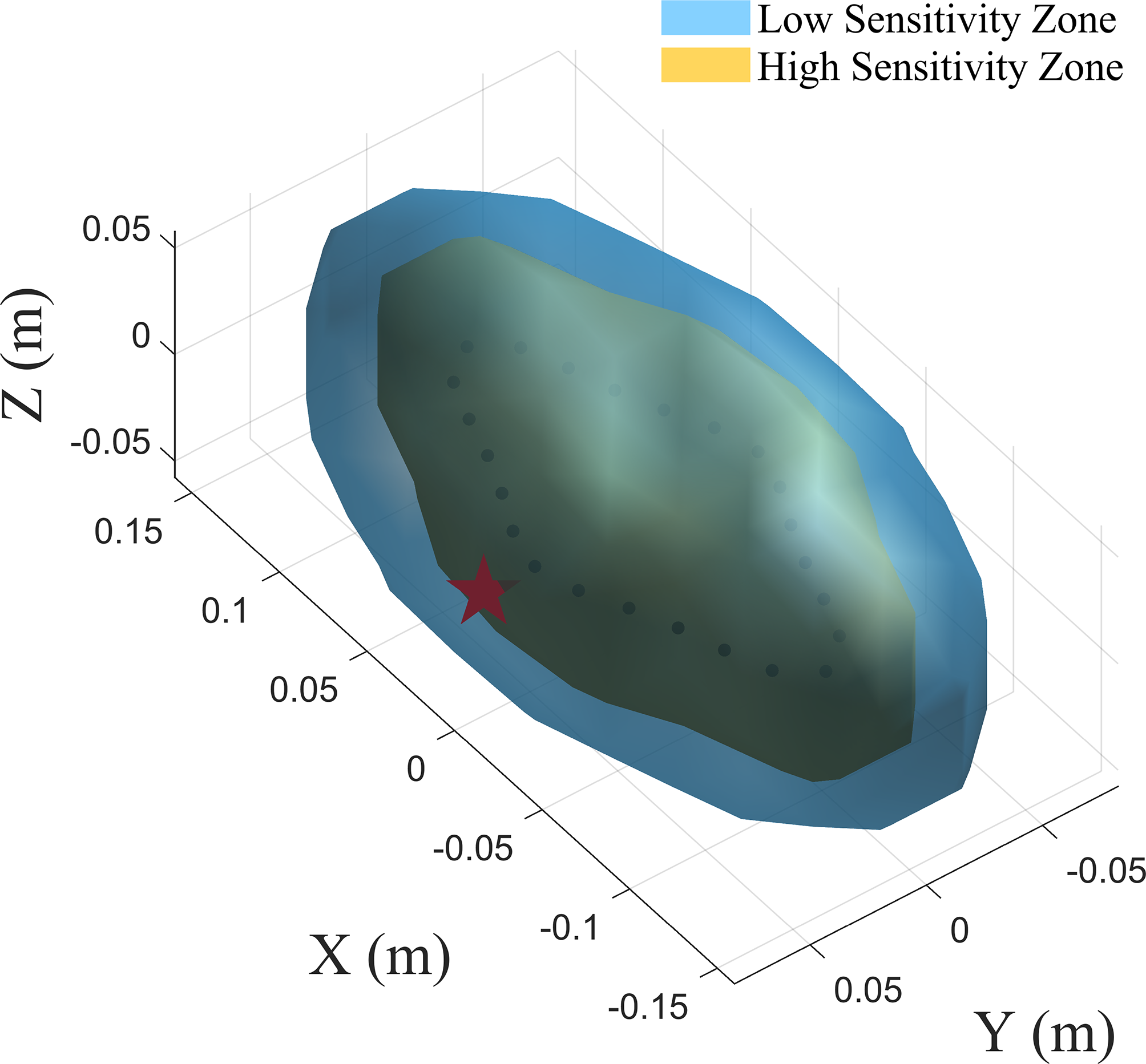}
   \caption{3D Array Sensitivity "Bubble"}
   \label{fig5_lb}
\end{figure}

\subsection{Spatial Coverage and Sensitivity Analysis}
To rigorously quantify the effective communication range and directional characteristics of the proposed system, we introduce a Volumetric Sensitivity Metric, denoted as $\mathcal{S}(\mathbf{r})$. This metric represents the total signal energy captured by the array from a potential source located at $\mathbf{r}$, which is mathematically equivalent to the Euclidean norm of the spatial steering vector:
\begin{equation}
    \mathcal{S}(\mathbf{r}) = \|\mathbf{h}(\mathbf{r})\| = \sqrt{\sum_{i=1}^{N} |G(\mathbf{r}, \mathbf{r}_i)|^2},
\end{equation}
where $G(\mathbf{r}, \mathbf{r}_i)$ is the dipole geometric factor for the $i$-th sensor. Fig. \ref{fig5_lb} visualizes the 3D isosurfaces of this sensitivity field, effectively mapping the "Communication Bubble" surrounding the AUV.The visualization reveals a distinct, layered spatial structure that defines the operational limits of the system. The inner core, depicted by the yellow isosurface (60\% maximum sensitivity), represents the High-Fidelity Zone where the array gain is sufficient to suppress severe turbulence and maximize SNR, ensuring a near-zero BER. This core is elongated along the longitudinal axis ($x$-axis), coinciding with the streamlined distribution of the lateral line sensors. Surrounding this core is the Effective Edge Zone, marked by the blue isosurface (30\% maximum sensitivity), which defines the boundary of the reliable link; beyond this shell, the aggregated signal energy drops below the decodability threshold of the BPSK demodulator. Unlike omnidirectional acoustic transducers, this "peanut-like" volumetric shape demonstrates significant Anisotropy, indicating that the system is most sensitive to neighbors positioned in the fore-and-aft sectors while exhibiting reduced sensitivity in the transverse directions. This directional selectivity provides a passive spatial filtering mechanism, naturally rejecting interference from non-cooperating entities located in the blind zones.

\begin{figure*}[thpb]
   \centering
   \includegraphics[width=1.2\columnwidth]{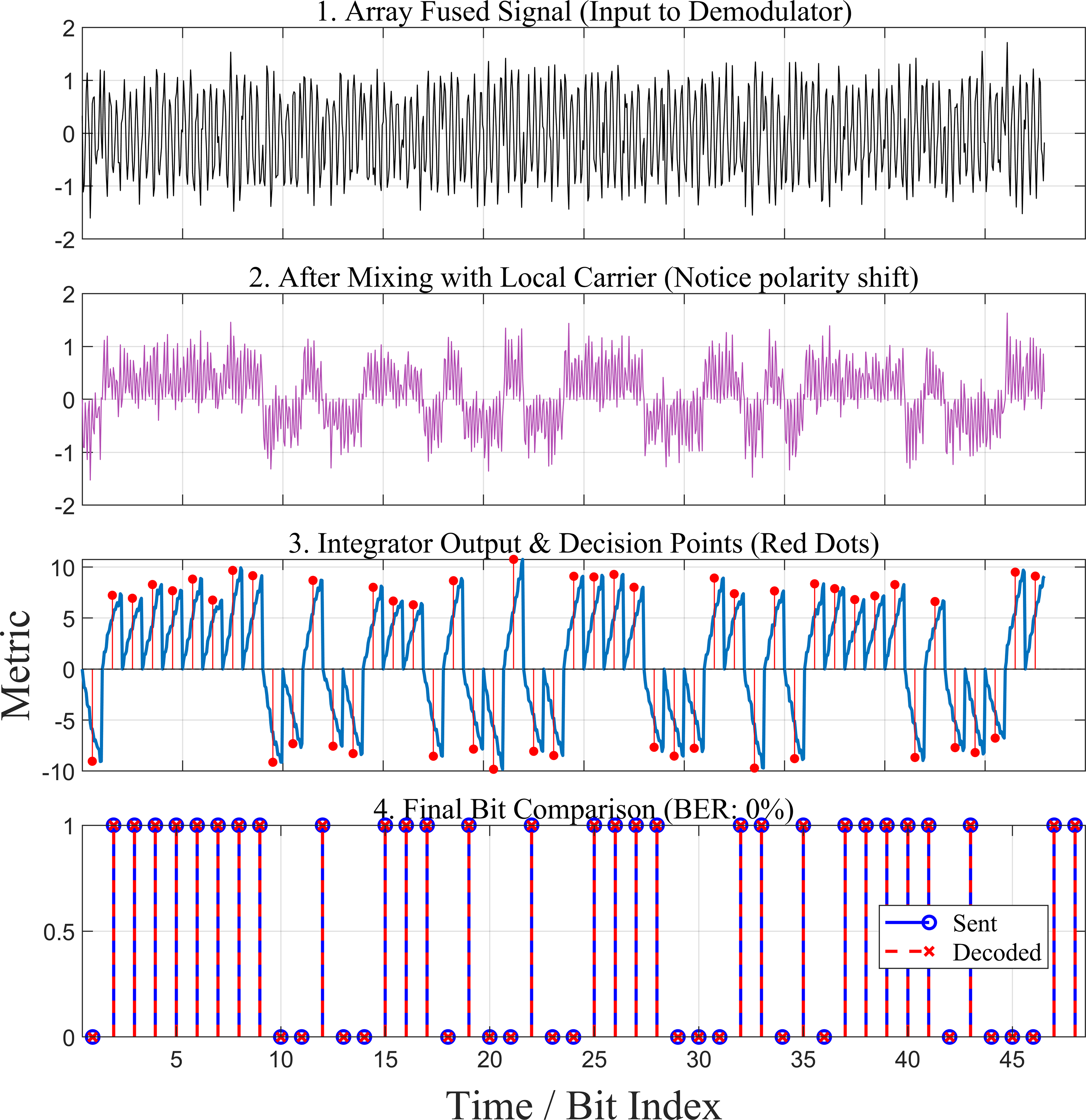}
   \caption{Demodulation Process}
   \label{fig5_lb}
\end{figure*}

\subsection{Temporal Demodulation Dynamics}
While spatial beamforming significantly enhances the SNR, the retrieval of digital information relies on robust temporal processing, the dynamics of which are delineated in Fig. \ref{fig5_lb}. The process initiates with the array-fused signal $y_{\Sigma}(t)$ (Panel 1), which, unlike the noisy raw sensor data, exhibits a distinguishable BPSK waveform structure where information is encoded in phase polarity. To extract this logic, the signal undergoes coherent mixing with a synchronized local carrier $c(t) = \sin(\omega_c t)$ (Panel 2); this operation effectively shifts the spectrum to baseband, acting as a "phase rectifier" that maps in-phase components (Bit '1') to positive amplitudes and anti-phase components (Bit '0') to negative ones. Subsequently, the Integrate-and-Dump filter (Panel 3) accumulates the coherent energy over each symbol period $T_s$, generating monotonic integration ramps while effectively averaging out zero-mean turbulence noise. The decision metrics $D_k$, marked by red dots at the sampling instants, exhibit a significant distance from the zero threshold, confirming a high noise margin. Ultimately, the perfect alignment between the decoded bits and the transmitted sequence (Panel 4) validates that the integration of spatial diversity with temporal coherence provides a robust decoding mechanism, achieving error-free transmission even in the presence of hydrodynamic disturbances.

\section{Discussion}

\subsection{Physical Limits of Mechanical Bandwidth}
 critical finding from our simulation is the intrinsic coupling between the carrier frequency ($f_c$) and the maximum achievable data rate ($R_b$). Unlike electromagnetic communications, where the carrier frequency ($\text{MHz} \sim \text{GHz}$) is orders of magnitude higher than the baseband signal, hydrodynamic communication is strictly constrained by the mechanical inertia of the actuator.We define the Cycle-per-Symbol Ratio (CSR) as $\gamma = f_c / R_b$ to quantify this constraint.In the successful regime where $R_b = 20 \, \text{bps}$ (corresponding to $\gamma = 2$), the system operates reliably because the coherent integrator accumulates energy over two full sinusoidal cycles, effectively averaging out orthogonal noise components. Conversely, simulations indicate that increasing the rate to $100 \, \text{bps}$ ($\gamma = 0.4$) results in immediate demodulation failure. Physically, this failure occurs because the actuator lacks the bandwidth to complete even a single vibration cycle within one bit period. This leads to severe ISI, where the phase state of the previous bit bleeds into the current integration window. Consequently, the capacity of a hydrodynamic channel is not primarily limited by the Shannon-Hartley theorem regarding SNR, but rather by the frequency response envelope of the mechanical system. This fundamental constraint necessitates a "Slow-Wave" design philosophy, prioritizing link reliability over raw throughput.

\subsection{The Stealth-Range Trade-off}

The proposed system operates in the near-field regime, where the pressure amplitude decays as $1/r^2$ (consistent with the dipole model derived in Section 2). While this rapid attenuation limits the effective communication range to approximately 10 body lengths ($< 1 \, \text{m}$), this limitation constitutes a significant tactical advantage for LPI operations.By contrast, traditional acoustic sonar signals decay as $1/r$ due to cylindrical or spherical spreading. A sonar signal strong enough to reach a neighbor 1m away may propagate residually for hundreds of meters, making it detectable by remote enemy hydrophones. The hydrodynamic dipole, with its steeper $1/r^2$ gradient, creates a sharply defined "Communication Bubble." Inside this bubble, the SNR is high due to the array gain; outside the bubble, the signal energy vanishes into the ambient turbulence floor almost instantly. This physical cutoff guarantees that the swarm's internal coordination remains invisible to remote surveillance, enabling truly silent tactical cooperation.

\section{Conclusion and Future Work}
\subsection{Conclusion}
This paper has presented "Hydrodynamic Whispering," a novel near-field silent communication paradigm tailored for AUV swarms. By repurposing the Artificial Lateral Line (ALL) from a passive sensor to an active communication interface, we have bridged the gap between biological inspiration and engineering application.The key contributions of this work are summarized as follows:Theoretical Foundation: We derived the dipole-based physical channel model, quantifying the distinct near-field propagation characteristics that enable stealth.System Architecture: A BPSK modulation scheme adapted for mechanical actuators was proposed, coupled with a 24-sensor conformal array receiver.Algorithmic Innovation: By introducing Spatial Matched-Field Beamforming, we demonstrated that a spatial diversity gain of $\approx 13.8 \text{dB}$ can be achieved, effectively recovering signals buried in strong turbulence noise (BER $\approx 0$ at SNR $< 0 \text{dB}$).This study confirms that hydrodynamic communication is not merely a biological curiosity but a viable engineering solution for the "last-meter" secure interaction in underwater robotics.

\subsection{Future Work}
While the simulation results are promising, transitioning from numerical models to practical deployment requires addressing significant physical and algorithmic challenges. Future research will prioritize experimental validation through the construction of a physical prototype utilizing a voice-coil vibration source and a MEMS pressure sensor array; this platform will be tested in a water flume to evaluate system resilience against real-world hydrodynamic disturbances, such as Von Kármán vortex streets. Concurrently, to address the Doppler frequency shifts induced by the relative motion in dynamic swarming scenarios, we plan to develop advanced Phase-Locked Loop (PLL) algorithms capable of real-time carrier tracking. Furthermore, we will investigate the feasibility of Full-Duplex operations, specifically exploring ntegrated ensing and ommunication frameworks. This would enable the Artificial Lateral Line to perform dual-function tasks, detecting environmental obstacles simultaneously while receiving data streams from peers, thereby maximizing the utility of the sensory hardware.

\addtolength{\textheight}{-12cm}   





\end{document}